\documentclass[a4paper,fleqn,usenatbib]{mnras}

\usepackage{newtxtext,newtxmath}

\usepackage[T1]{fontenc}
\usepackage{ae,aecompl}

\usepackage{graphicx}	
\usepackage{amsmath}	
\usepackage{amssymb}	
\usepackage{soul}       

\title[]{Role of gaseous giants in the dynamical evolution of terrestrial planets and water delivery in the habitable zone} 
\author[M. B. S\'anchez et al.]{
Mariana B. Sanchez$^{1,2}$,\thanks{E-mail: msanchez@fcaglp.unlp.edu.ar}
Gonzalo C. de El\'ia$^{1,2},$
and Luciano A. Darriba$^{1,2}$
\\
$^{1}$Facultad de Ciencias Astron\'omicas y Geof\'isicas, Universidad Nacional de La Plata\\ Paseo del Bosque s/n, La Plata, B1900FWA, Buenos Aires, Argentina
\\
$^{2}$Instituto de Astrof\'isica de La Plata, CCT La Plata-CONICET-UNLP\\ Paseo del Bosque s/n, La Plata, B1900FWA, Buenos Aires, Argentina
}

\date{Accepted XXX. Received YYY; in original form ZZZ}

\pubyear{2018}

\begin{document}
\label{firstpage}
\pagerange{\pageref{firstpage}--\pageref{lastpage}}
\maketitle

\begin{abstract}
In the present research, we study the effects of a single giant planet in the dynamical evolution of water-rich embryos and planetesimals, located beyond the snow line of systems around Sun-like stars, in order to determine what kind of terrestrial-like planets could be formed in the habitable zone (hereafter HZ) of these systems. To do this, we carry out $N$-body simulations of planetary accretion, considering that the gas has been already dissipated from the disk and a single giant planet has been formed beyond the snow line of the system, at 3 au. We find that a giant planet with a value of mass between Saturn-mass and Jupiter-mass, represents a limit from which the amount of water-rich embryos that moves inward from beyond the snow line starts to decrease. From this, our research suggests that giant planets more massive than one Jupiter-mass become efficient dynamical barriers to inward-migrating water-rich embryos. Moreover, we infer that the number of these embryos that survive in the HZ significantly decreases for systems that host a giant planet more massive than one Jupiter-mass. This result has important consequences concerning the formation of terrestrial-like planets in the HZ with very high water contents and could provide a selection criteria in the search of potentially habitable exoplanets in systems that host a gaseous giant around solar-type stars.
\end{abstract}




\section{Introduction}

At the present time, we know that planetary systems are common in the Universe. They could be found around different stars and be composed by all kind of planets with a huge variety of parameters. Up to date, there are 3824 confirmed exoplanets and 2859 planetary systems \texttt{(http://exoplanet.eu/)}, and more objects are waiting to be confirmed. As the years pass, the observational techniques improve and the theoretical models become more refined. In fact, observational studies as \citet{Cumming2008} and \citet{Howard2013}, and theoretical works as \citet{Mordasini2009}, \citet{Ida2013}, and \citet{Ronco2017} have shown the existence of a wide diversity of planetary systems, suggesting that those systems consisting only of rocky planets would seem to be the most common in the Universe.

Of particular interest are the terrestrial-like planets located in the so-called ``habitable zone'' (hereafter HZ) of a given system, which is defined as the circumstellar region inside which a planet can retain liquid water on its surface. However, the location of a terrestrial-like planet in the HZ is a necessary condition but not enough to say that such a planet could host life as we know today. In fact, the maintenance of habitable conditions on a
planet requires to satisfy other conditions, which are related to the existence of a suitable atmosphere, organic material, the presence of a magnetic field, plate tectonics that replenish the atmosphere of CO2, among others \citep{Martin2006}. 

Several authors worked with $N$-body simulations in order to describe the possible formation and evolution of a planetary system and the water delivery in the HZ in different dynamical scenarios. In particular, many works focused on the study of planetary systems that host at least one gaseous giant. For example, \citet{Raymond2004} and \citet{Raymond2006} explored the accretion process and dynamics of terrestrial planets around a Sun-like star under the effects of a Jovian planet in the outer disk, while \citet{Mandell2007} studied the formation of Earth-like planets during and after giant planet migration in solar-type stars, considering systems with a single migrating giant planet and other ones with an inner migrating gas giant and an outer non-migrating giant planet. Moreover, \citet{Raymond2011} studied the terrestrial-like planet formation and water delivery in systems with multiple unstable gas giant planets. On the other hand, \citet{Raymond2007} studied the habitable planet formation considering one Jovian-type planet around binary systems. They worked with a Sun-like star as the primary star and took values of 0.5 $\textrm{M}_{\textrm{\sun}}$, 1 $\textrm{M}_{\textrm{\sun}}$, and 1.5 $\textrm{M}_{\textrm{\sun}}$ for the secondary star of the binary system, focusing their attention on the formation of Earth-like planets in the HZ of the primary star. More recently, \citet{Quintana2014} worked with terrestrial planet formation around a Sun-like star considering a massive planet in the system with values of masses between 1 $\textrm{M}_{\textrm{\earth}}$ and 1 $\textrm{M}_{\textrm{jup}}$, while \citet{Izidoro2015} used dynamical simulations to show that gas giant planets act as barriers to the inward migration of super-Earths initially located in distant orbits around a Sun-like star, considering interactions with a gaseous protoplanetary disk in the middle stages of their formation. Lately, \citet{Zain2018} worked with planetary formation and water delivery in the HZ around a Sun-like star considering different scenarios: one with a Jovian-like giant, one with a Saturn-like giant and other ones without giant planets in the system. They found that planets with high amount of water in mass, were formed in the HZ of all their work scenarios. All these works focus their attention in the late stages of terrestrial planet formation, assumed that water was delivered to planets via collisions, considering the condensation of material beyond the snow line, located about 3 \textrm{au}.

In the present work, we use $N$-body simulations in order to study the dynamical evolution of systems that host a massive gaseous giant just beyond the snow line around a Sun-like star, when the gas has been already dissipated of the disk. The main goal of our research is to understand how the giant planet of a system affects the formation of the terrestrial ones, in particular those potentially habitable. To do this, we propose different scenarios, considering only one giant planet per system around the snow line, whose mass ranges from 0.5 $\textrm{M}_{\textrm{sat}}$ to 3 $\textrm{M}_{\textrm{jup}}$, where $\textrm{M}_{\textrm{sat}}$ and $\textrm{M}_{\textrm{jup}}$ represent the planetary mass of Saturn and Jupiter, respectively. This work is structured as follows: in Section 2, we describe the model and the numerical method that we used for selecting the initial conditions of our work. In Section 3, we present the $N$-body code and specify the physical and orbital parameters of the bodies that participate of the numerical simulations. In Section 4, we show the HZ model that we used in order to classify the potentially habitable planets. In Section 5, we expose the results obtained from the $N$-body simulations. At the end, we give the conclusions of the present research in Section 6.

\section{Model and numerical method}

In this section, we describe the model used for the protoplanetary disk together with the parameters chosen for generating the initial conditions of our numerical simulations. From these initial conditions, we calculate the distribution of embryos, planetesimals, and the location of the giant planet at the beginning of the post gas phase of each system, in order to carry out the $N$-body simulations on this last stage of formation of a planetary system.

\subsection{Model of the protoplanetary disk}

The parameter that determines the distribution of the material in a protoplanetary disk is the surface density. The gas-surface density profile $\Sigma_{\textrm{g}}(R)$ and the solid-surface density profile $\Sigma_{\textrm{s}}(R)$  that we adopted for our model of protoplanetary disk are given by
\begin{equation}
\Sigma_{\textrm{g}}(R)=\Sigma_{0\textrm{g}}\left(\frac{R}{R_{\textrm{c}}}\right)^{-\gamma}\textrm{e}^{{-(R/R_{\textrm{c}})}^{2-\gamma}},
\label{eq:densgas}
\end{equation}
\begin{equation}
\Sigma_{\textrm{s}}(R)=\Sigma_{0\textrm{s}}\eta_{\textrm{ice}}\left(\frac{R}{R_{\textrm{c}}}\right)^{-\gamma}\textrm{e}^{{-(R/R_{\textrm{c}})}^{2-\gamma}},
\label{eq:denssol}
\end{equation}
\citep{Lynden-Bell1974,Hartmann1998} where $R$ is the radial coordinate in the mid plane of the protoplanetary disk, $R_{\textrm{c}}$ the characteristic radius of the disk, and $\gamma$ the factor which determines the density gradient. Moreover, the parameter $\eta_{\textrm{ice}}$ represents an increase in the amount of solid material due to the condensation of water beyond the snow line $ R_{\textrm{ice}}$.
The normalization constant $\Sigma_{0\textrm{g}}$ is determined assuming axial symmetry for the material of the disk. Under such conditions, we can express the protoplanetary disk mass by
\begin{equation}
  M_{\textrm{d}}=\displaystyle\int_{0}^{\infty} 2\pi R\Sigma_{\textrm{g}}(R)\textrm{d}R,
\label{eq:mdisco}
\end{equation}
from which we obtain
\begin{equation}
\Sigma_{0\textrm{g}} = (2 - \gamma) \frac{M_{\textrm{d}}}{2\pi R_{\textrm{c}}^{2}}.
\end{equation}
In order to determine $\Sigma_{0\textrm{s}}$ we used the relation between the gas and solid surface densities given by \citet{Lodders2009}. As we are considering a 1 M$_{\odot}$ star with solar metallicity, this relation is confined to $\Sigma_{0\textrm{s}}=z_{0}\Sigma_{0\textrm{g}}$, where $z_{0}$ is the primordial abundance of heavy elements in the Sun and has a value of $z_{0}=0.0153$.

It is worth remarking that the systems of work that we propose in the present research host a giant planet around the snow line at the end of the gaseous phase. Thus, we must select appropriate parameters concerning the disk mass and the gas and solid density profiles that lead to the formation of such work scenarios. To do this and following the study developed by \citet{Zain2018}, we adopt a disk mass $M_{\textrm{d}} = 0.1$ $\textrm{M}_{\odot}$, which gives the amount of material necessary around the snow line to form a gaseous giant planet in the range of masses we are working with. Despite of the existence of observed less massive protoplanetary disks, \citet{Andrews2010} found evidence of more massive protoplanetary disks in the 1 Myr-old Ophiuchus star-forming region $\geq10\%$ of the stellar-mass, such as Elias 24 and $DoAr$ $25$ with $11.7\%$ and $13.6\%$ of the stellar-mass respectively. We used a characteristic radius $R_{\textrm{c}} = 25$ $\textrm{au}$, and an exponent $\gamma = 0.9$, which are consistent with the observations carried by \citet{Andrews2010}. According to \citet{Lodders2009}, we assume $\eta_{\textrm{ice}}=0.5$ if $R < R_{\textrm{ice}}$ and $\eta_{\textrm{ice}}=1$ if $ R > R_{\textrm{ice}}$, being $R_{\textrm{ice}}=2.7$ $\textrm{au}$ for a solar luminosity star \citep{Ida2004}. Furthermore, we assume that the protoplanetary disk presents a radial compositional gradient. In fact, we consider that bodies beyond $R_{\textrm{ice}}$ present a water content of 50\% by mass, while bodies inside $R_{\textrm{ice}}$ do not have water. This water distribution is assigned to each body in our simulations based on its initial location. 
 
\subsection{Post gas stage: initial distributions}

From the surface density profiles specified in the previous section, we determine the initial position of the giant planet and the embryo and planetesimal distributions in the post gas stage of the system. We remark that the region of study of the present research is confined between $0.5$ $\textrm{au} \leq R \leq 9.5$ $\textrm{au}$, and it includes the HZ of the system, the snow line, and an outer region with water-rich embryos and planetesimals. To give a better understanding of the different body distributions, we divide the region of work in the following three parts:
\begin{enumerate}
\item {\it Inner region}, with $0.5$ $ \textrm{au} \leq R < 2.5$ $\textrm{au}$,
\item {\it Central region}, with $2.5$ $\textrm{au} \leq R < 3.5$ $\textrm{au}$,
\item {\it Outer region}, with $3.5$ $\textrm{au} \leq R \leq 9.5$ $\textrm{au}$.
\end{enumerate}
In the following, we describe the considerations adopted to determine the distributions of the different bodies in each of these three regions.

\subsubsection{Inner region}

In the inner region of the system, we consider only the existence of planetary embryos since we assume that all planetesimals were already accreted by embryos in the previous stages such as it was shown by \citet{Zain2018}. For a suitable inner embryo distribution, we express the mass of an embryo located at $R$ growing in the oligarchic growth mode by

\begin{equation}
M=2\pi R\Delta R_{\textrm{H}}\Sigma_{\textrm{s}}(R)f,
\label{eq:masaemb}
\end{equation}
\citep{Kokubo2000} where $\Sigma_{\textrm{s}}(R)$ is the solid-surface density, $f$ a factor that represents the planetesimal fraction accreted by the embryo, and $\Delta R_{\textrm{H}}$ the orbital separation between two consecutive embryos of mass $M$ in terms of their mutual Hill radii, which is given by
\begin{equation}
R_{\textrm{H}}=R\left(\frac{2M}{3 M_{\star}}\right)^{\frac{1}{3}},
\label{eq:radiohill}
\end{equation}
being $M_{\star}$ the mass of the central star. Replacing Eqs. \eqref{eq:denssol} and \eqref{eq:radiohill} in Eq. \eqref{eq:masaemb}, we obtain an expression for the mass of each embryo as a function of the distance $R$, which is given by
\begin{equation}
M=\left(2 \pi R^{2} \Delta \Sigma_{0\textrm{s}} \eta_{\textrm{ice}} f \left(\frac{2}{3 M_{\star}}\right)^{\frac{1}{3}}\left(\frac{R}{R_{\textrm{c}}}\right)^{-\gamma}\textrm{e}^{{-\left(\frac{R}{R_{\textrm{c}}}\right)}^{2-\gamma}}\right)^{\frac{3}{2}}.
\end{equation}
In the inner system, we assume that $f=1$, since it is assumed that all planetesimals were accreted by the inner embryos at the end of the gaseous phase \citep{Zain2018}. For the initial mass of the first embryo, which is located at $R_0 =$ 0.5 $\textrm{au}$, we derive a value of $M_{0} =$ 0.11 $\textrm{M}_{\oplus}$. Assuming a $\Delta =$ 5, we calculate the initial locations and masses for the rest of the inner embryos by the expressions

\begin{equation}
  R_{i+1} = R_{i} + \Delta R_{i} \left(\frac{2M_{i}}{3M_{\star}}\right)^{\frac{1}{3}},
  \label{eq:distancias}
\end{equation}
\begin{equation}
  M_{i+1}= \left(A\left(\frac{2}{3 M_{\star}}\right)^{\frac{1}{3}}\left(\frac{R_{i+1}}{R_{\textrm{c}}}\right)^{-\gamma}\textrm{e}^{{-\left(\frac{R_{i+1}}{R_{\textrm{c}}}\right)}^{2-\gamma}}\right)^{\frac{3}{2}},
  \label{eq:masas} 
\end{equation}
for $i =$ 0, 1, $\dots$, being $A=2 \pi R^{2} \Delta \Sigma_{0\textrm{s}} \eta_{\textrm{ice}}f$. By doing this up to a value of $R <$ 2.5 au, we obtain a population of 37 inner embryos, which has a total mass of $16.78$ $\textrm{M}_{\oplus}$.

\subsubsection{Central region}
\label{sec:centralregion}

In the central region of the system, we only located the giant planet at $R = 3$ $\textrm{au}$, just beyond the snow line, as we assumed that it had already accreted all the surrounding material in the previous stages. We considered that in an environment around the snow line is where the surface density is maximized and the influence zone is suitable for an embryo to accrete enough mass in order to form a giant core \citep{Brunini2008}. According to a core instability model and an oligarchic growth regime of solid protoplanets for the growth of a core of a giant planet, the expected necessary mass is one between 10 $\textrm{M}_{\oplus}$ - 15 $\textrm{M}_{\oplus}$, \citep{Mizuno1980,Bodenheimer1986,Pollack1996}, which is the amount of mass of the surrounding material, taking into account embryos and planetesimals, which were located in this region in the previous stages if we consider a disk mass of 0.1 $\textrm{M}_{\odot}$ \citep{Zain2018}. Assuming also that the giant planet was formed $in$ $situ$, we defined each work scenario in relation with the giant planet mass chosen by:

\begin{description}
\item [-] \textbf{Scenario 1}: $M_{\text{giant}} = 0.5 \textrm{M}_{\textrm{sat}}$.
\item [-] \textbf{Scenario 2}: $M_{\text{giant}} = 1 \textrm{M}_{\textrm{sat}}$.
\item [-] \textbf{Scenario 3}: $M_{\text{giant}} = 1 \textrm{M}_{\textrm{jup}}$.
\item [-] \textbf{Scenario 4}: $M_{\text{giant}} = 1.5 \textrm{M}_{\textrm{jup}}$.
\item [-] \textbf{Scenario 5}: $M_{\text{giant}} = 2 \textrm{M}_{\textrm{jup}}$.
\item [-] \textbf{Scenario 6}: $M_{\text{giant}} = 3 \textrm{M}_{\textrm{jup}}$.
\end{description}

We chose masses between $0.5\textrm{M}_{\textrm{sat}}$ and $3\textrm{M}_{\textrm{jup}}$ because we wanted to study massive giant planets effects as barriers of water-rich material which moves inward the inner system and their relation with the amount of water in planets into the HZ. We chose a range in mass which includes the mass of the gaseous giant planets of our Solar System, considering an arbitrary lower limit of a half of the Saturn-Mass and an upper limit of 3 times the Jupiter-mass, to analyze the effects of little more massive giant planets than the massive one in our Solar-System (less massive planets formed around the snow line of a system which orbits around a Sun-like star have been studied in \citet{Zain2018}).

\subsubsection{Outer region}
  
\begin{figure}
\includegraphics[width=0.45\textwidth]{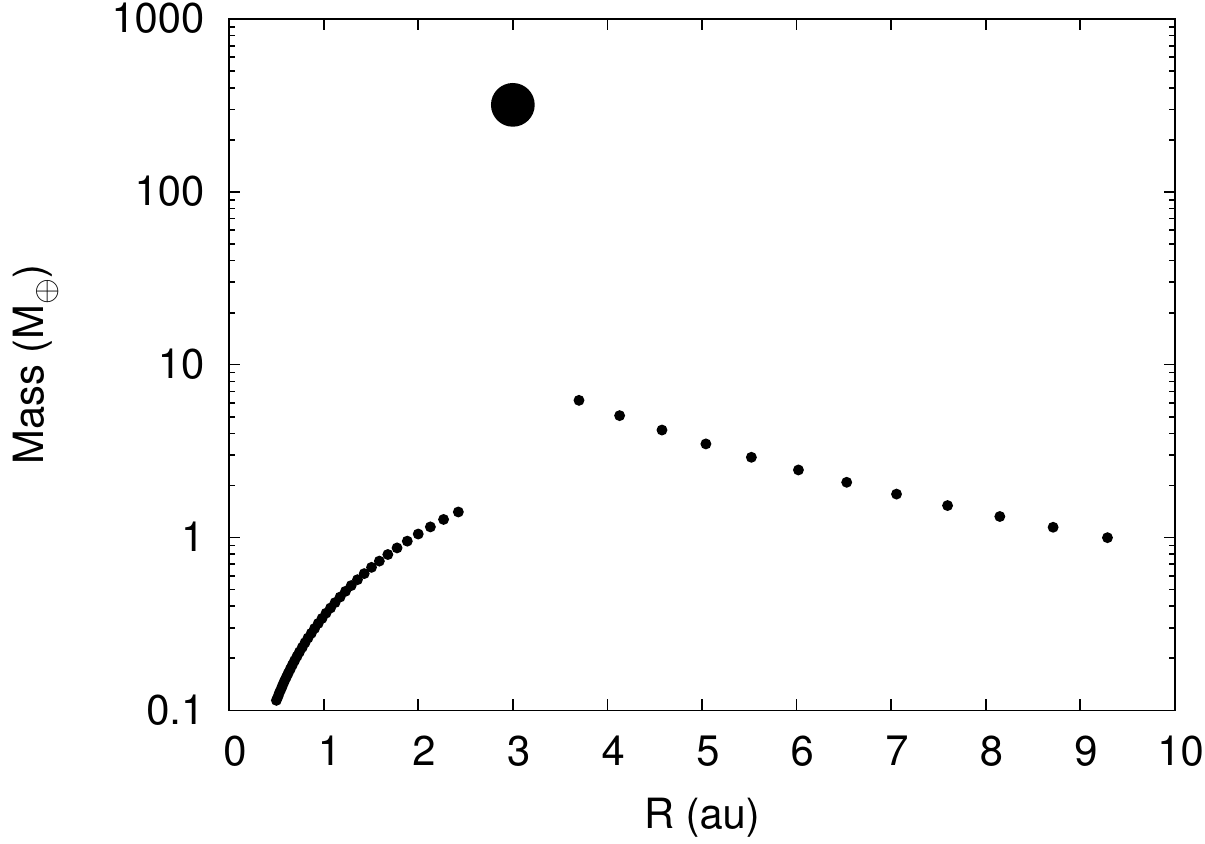}
\caption[Embryo distribution]{Initial distribution of embryo masses and the giant planet of 1 $\textrm{M}_{\textrm{jup}}$ as a function of the distance $R$ from the central star.}
\label{fig:embriones}
\end{figure}

In the outer region of the system, we distribute both embryos and planetesimals. As for the embryo distribution, we consider that the factor $f$ is a function of the radial distance $R$ from the central star, which is constructed in a consistent way with the outer embryo distribution observed in \citet{Zain2018}. Then, we use the Eq. \eqref{eq:distancias} to determine the initial locations of the outer embryos and the Eq. \eqref{eq:masas} to calculate their masses. From this, we obtain a total mass of 33.22 $\textrm{M}_{\oplus}$, which is distributed in 12 outer embryos. In Fig.~\ref{fig:embriones}, we can see the variation of the mass of each embryo and the giant planet of 1 $\textrm{M}_{\textrm{jup}}$ as a representative giant, as a function of the distance $R$.

\begin{figure}
  \includegraphics[angle=270,width=0.45\textwidth]{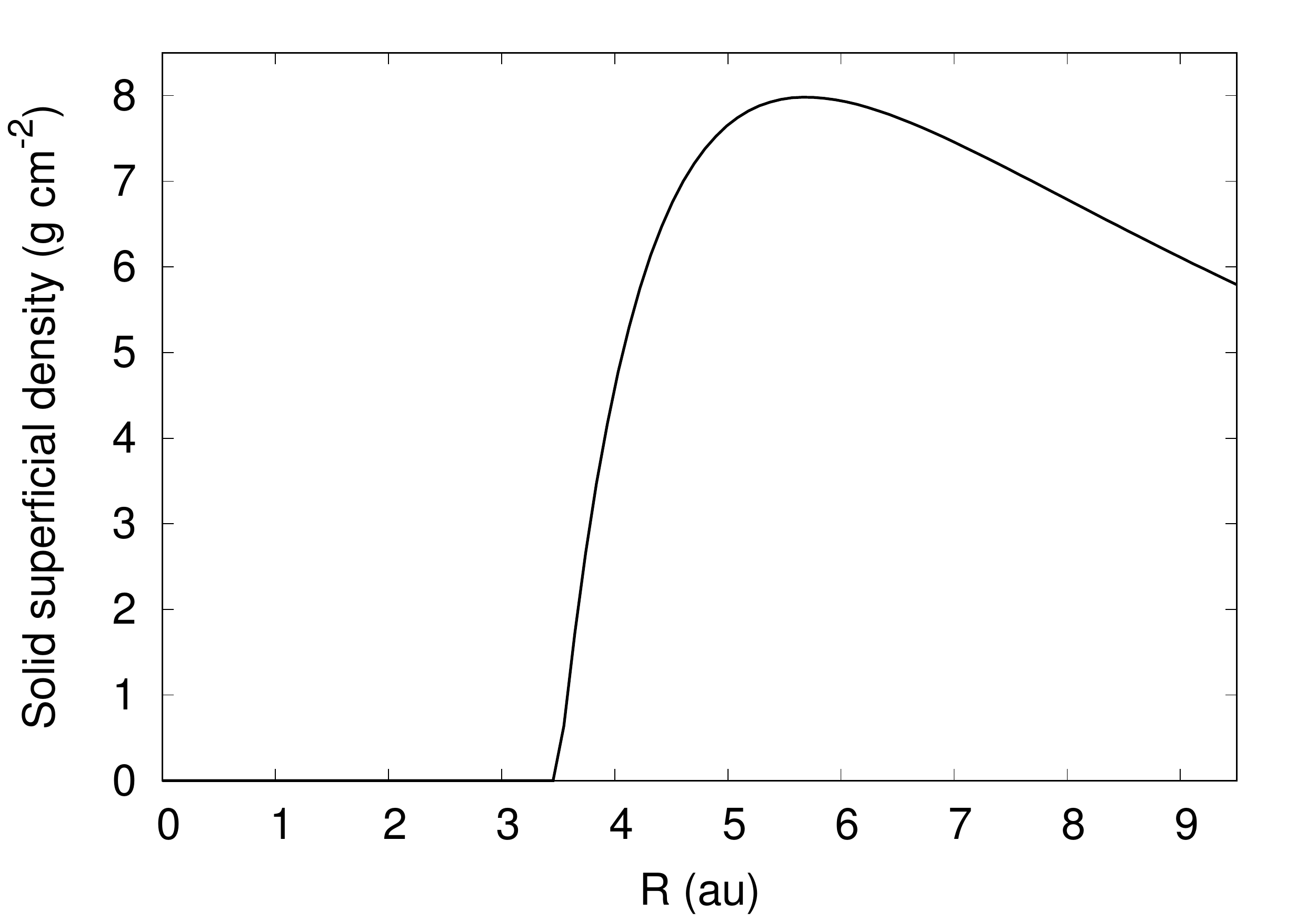}
\caption[Planetesimals distribution]{Solid initial density profile which models the planetesimals distribution at the beginning of the post gas stage as a function of the distance $R$.}
\label{fig:planetesimales}
\end{figure}

Regarding planetesimals, we compute the total mass contained in such a population from the difference between the total mass of solids in the outer region and the total mass contained in the outer embryo population. From this, the total mass of the planetesimal population associated to the outer region of the system is equal to 55 $\textrm{M}_{\oplus}$. To analyze the planetesimal distribution, we multiply the solid-surface density profile given by Eq. \eqref{eq:denssol} for $1-f$. As we have already mentioned, $f = 1$ in the inner region, while $f$ is a function of the radial distance $R$ in the outer region, which is constructed in a consistent way with the outer embryo distribution observed in \citet{Zain2018}. From this, we calculate the planetesimal surface density profile at the beginning of the post gas stage, which is shown in Fig.~\ref{fig:planetesimales}. As the reader can see, the planetesimal surface density profile is zero in the inner region of the system, since we assume that the embryos efficiently accreted all planetesimals of such a region during the gaseous phase.

According to this, we can express the differential mass of planetesimals contained in a ring centered at $R$ with a width d$R$ by
\begin{equation}
\textrm{d}M=2 \pi R\Sigma_{\textrm{s}}(R)(1-f)\textrm{d}R.
\label{eq:difmasapl}
\end{equation}
If we consider that all the planetesimals have the same individual mass, we can write the mass $\textrm{d}M$ as a function of the individual mass of a planetesimal $m_{\textrm{p}}$ and the amount of planetesimals $\textrm{d}N$ in the ring by 
\begin{equation}
\textrm{d}M=m_{\textrm{p}}\textrm{d}N.
\label{eq:pl}
\end{equation} 
Then, we sample the amount of planetesimals with a distribution function $F(R)$ that is function of the distance $R$ from the central star
\begin{equation}
\textrm{d}N=F(R)\textrm{d}R.
\label{eq:dn}
\end{equation}
Finally, from the Eqs. \eqref{eq:difmasapl}, \eqref{eq:pl}, and \eqref{eq:dn}, we obtain the following expression for the distribution function $F(R)$ 
\begin{equation}
F(R)=\frac{2\pi R}{m_{\textrm{p}}}\Sigma_{\textrm{s}}(R)(1-f).
\label{eq:funcdist}
\end{equation} 
This function $F(R)$ will be used to generate the planetesimal population for developing our $N$-body simulations. 

\section{$N$-\textrm{body} simulations}

To carry out our study, we use the so-called $N$-body code \textsc{Mercury}, which was developed by \citet{Chambers1999}. In our work, we make use of the hybrid integrator, which uses a second order symplectic algorithm to treat interaction between objects with separations greater than 3 Hill radii and a Bulirsch-st\"oer method for resolving closer encounters. We integrate each simulation for 100 Myr, which is consistent with the Earth-formation timescale according to \citet{Jacobson2014}. To develop the integration, we adopted a time step of 3 days, which is shorter than $1/40$th of the orbital period of the innermost body in our simulation. Moreover, in order to avoid any numerical error for small-perihelion orbits, we assume a non-realistic Sun's radius of 0.1 \textrm{au}.

\textsc{Mercury} evolves the orbits of embryos and planetesimals allowing collisions between them. These collisions are treated as perfectly inelastic ones, conserving the mass and water content of the bodies of the simulations. In order to reduce the CPU time, we assume that embryos interact gravitationally with all other bodies of the simulation, while planetesimals are not self-interacting \citep{Raymond2006}.

To make use of the \textsc{Mercury} code, it is necessary to give physical and orbital parameters for the giant planet, embryos and planetesimals of our simulations. Regarding giant planet, we change its mass in each work scenario, adopting values of  0.5 \textrm{M}$_{\textrm{sat}}$, 1 \textrm{M}$_{\textrm{sat}}$, 1 \textrm{M}$_{\textrm{jup}}$, 1.5 \textrm{M}$_{\textrm{jup}}$, 2 \textrm{M}$_{\textrm{jup}}$ and 3 \textrm{M}$_{\textrm{jup}}$, such as it was proposed in Sect. 2.2.2. For the most massive giant planets (1 \textrm{M}$_{\textrm{jup}}$, 1.5 \textrm{M}$_{\textrm{jup}}$, 2 \textrm{M}$_{\textrm{jup}}$ and 3 \textrm{M}$_{\textrm{jup}}$), we consider a physical density of 1.3 g cm$^{-3}$, while for the less massive giants (0.5 $\textrm{M}_{\textrm{sat}}$ and 1 $\textrm{M}_{\textrm{sat}}$) we assume a physical density of 0.7 g cm$^{-3}$. Moreover, each giant planet has an initial semi-major axis of $a = 3$ $\textrm{au}$. Finally, we consider initial quasi-circular and co planar orbits for every giant planet, with values of perihelion argument $\omega$, ascending node longitude $\Omega$, and mean anomaly $M$ randomly selected between $0^{\circ}$ and $360^{\circ}$.

In all our work scenarios, we consider the same distribution of masses and semi-major axis for the planetary embryos. In fact, each of them is assigned with an initial semi-major axis and an initial mass given by Eqs. \eqref{eq:distancias} and \eqref{eq:masas}, respectively. For all embryos, we consider a physical density of 3 g cm$^{-3}$. Moreover, initial eccentricities and inclinations lower than 0.02 and $0.5^{\circ}$, respectively, are randomly assigned, while the initial angular parameters $\omega$, $\Omega$ and $M$ are randomly determined between $0^{\circ}$ and $360^{\circ}$ for each embryo. 

In all our simulations, we include 1000 planetesimals with an individual mass of 0.055 $\textrm{M}_{\oplus}$. To determine the initial values of the semi-major axis of the planetesimals, we adopt the acceptance-rejection method developed by John von Neumann using the function $F(R)$ given by Eq. \eqref{eq:funcdist}. For all planetesimals, we consider a physical density of 1.5 g cm$^{-3}$. Finally, just like for embryos, the planetesimals are randomly assigned with initial eccentricities and inclinations lower than 0.02 and $0.5^{\circ}$, respectively, while $\omega$, $\Omega$ and $M$ are randomly given between $0^{\circ}$ and $360^{\circ}$.   

Because of the stochastic nature of the accretion process, we remark that it is very necessary to carry out a set of $N$-body simulations for each of our 6 work scenarios in order to analyze the results in a statistical way. Thus, for scenarios 2-6, we carry out 30 numerical simulations, while that for scenario 1 we develop only 13 simulations because they require much more CPU time. It is worth noting that the energy is conserved better than 1 part in 10$^{4}$ in all cases.

\section{Habitable zone model}
\label{sec:hzmodel}

In the present work, we use the HZ definition given by \citet{Kopparapu2013}, \citet{Kopparapu2013e} and \citet{Kopparapu2014}. Using an updated 1-D radiative-convective, cloud free model, they estimated the boundaries of the HZ around Sun-like stars. In their studies, they took new H$_{2}$O and CO$_{2}$ absorption coefficients, derived from the HITRAN 2008 and HITEMP 2010 line-by-line databases. In this model, they supposed that into the inner HZ, atmospheres are dominated by water, while in the outer HZ, are dominated by carbon dioxide. Moreover, considering these kind of atmospheres and the relation between the pressure of surrounding nitrogen and the planetary radius, they found a dependence between the planetary mass and the width of the HZ. In fact, they studied the limits of the conservative HZ for planets of 0.1 M$_{\oplus}$, 1 M$_{\oplus}$ and 5 M$_{\oplus}$ which orbit around a Sun-like star, founding a changing internal limit of 1.005 $\textrm{au}$, 0.9504 $\textrm{au}$ and 0.9174 $\textrm{au}$ respectively. They also found a fix external limit of 1.676 $\textrm{au}$ in all cases. We could notice that the internal limit gets closer to the star while the planetary mass increases what concludes that with a more massive planet a wider HZ is related. From the values of the internal limit in relation with the planetary mass, we did an interpolation in order to associated our planets with their limits of the HZ. 

In our work, we considered that a planet with its semi-major axis, aphelion and perihelion completely contained into the HZ, will be a potentially habitable planet. However, in our simulations, we found some planets with high eccentricities. In case that a planet has a very eccentric orbit and its perihelion or aphelion scape from the limits of the HZ, \citet{Williams2002} proposed that is the average temporal flux in an orbit which determines the habitability conditions. We adopted this average flux criterion when the perihelion or aphelion of a planet is not completely contained into the limits of the HZ but near them. For a planet with eccentricity $e$, semi-major axis $a$, and assuming a Sun-like central star, the average flux normalized to terrestrial flux is
\begin{equation}
S_{\textrm{eff}}=\frac{1}{a^{2}\sqrt{1-e^{2}}}.     
\label{eq:flujo}
\end{equation}
On the other side, \cite{Kopparapu2014} calculated values of the average flux for 0.1 M$_{\oplus}$, 1 M$_{\oplus}$ and 5 M$_{\oplus}$ founding changing maximum average fluxes of 0.99, 1.107 and 1.188 respectively, which are normalized to terrestrial flux. They also found a fix minimum average flux of 0.356, normalized to terrestrial flux, in all cases. We could noticed that the maximum flux gets bigger while the planetary mass increases. We used those maximum average flux values in an interpolation with their planetary mass in order to calculate maximum average fluxes for our planets and related them with the orbital elements semi-major axis and eccentricity through the Eq. \eqref{eq:flujo}, in order to determinate flux curves of maximum average flux that allow us to express the semi-major axis as a function of the eccentricity. Using that flux criterion, we can say that a planet is considered inside the HZ if the evolution of its semi-major axis and eccentricity, in the last period of the integration, are contained into the maximum and minimum average flux curves.\\
In this context, we assumed both criterion to consider a planet as a potentially habitable one. We will see this classification in detail in section Results.\\

\section{Results}

Once the work scenarios were defined, our main goal is to analyze the influence of a single giant planet located around the snow line in the dynamical evolution of each planetary system of our simulations. In particular, we study how a single giant planet affects the evolution of water-rich embryos and planetesimals located beyond the snow line, in order to understand what kind of planets could be formed in the HZ of the system and their amount of water in mass in each of our six work scenarios.

\subsection{Dynamical evolution of outer embryos}
\label{sec:embryoevol}

\begin{figure}
\includegraphics[width=0.45\textwidth]{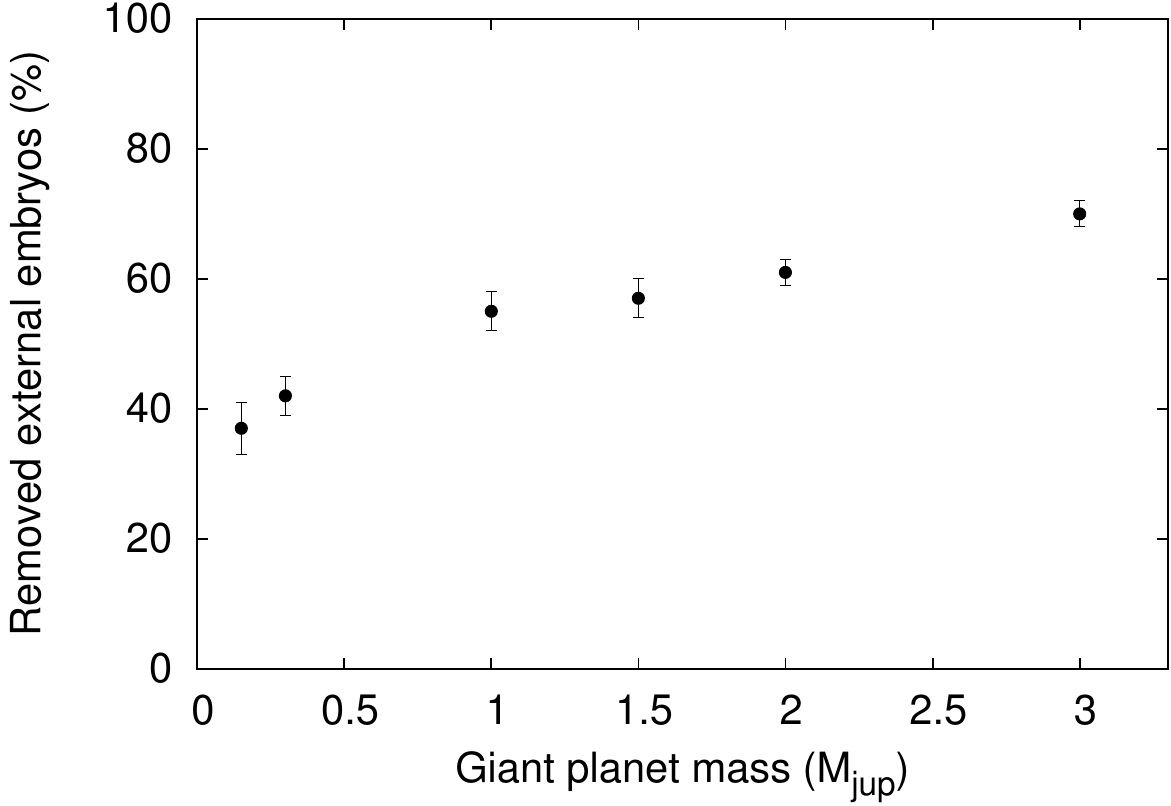}
\caption[Removed outer embryos]{Average percentage of the amount of removed outer embryos in each work scenario with their associated error.}
\label{fig:remocion}
\end{figure}

First of all, we study how a single giant planet located around the snow line of a system is able to affect dynamical evolution of water-rich embryos and planetesimals of the outer disk, which should have important implications in the type of planets formed in the HZ of the system in each work scenario. To study the evolution of the outer embryo population, we started analyzing the removal process of outer embryos in each scenario in order to determine the percentage of removed outer embryos as a function of the giant's mass. To do this, we calculated the number of removed outer embryos in every simulation for each scenario of study. Then, we computed averaged percentage adding the number of removed outer embryos in each simulation and normalizing it to the total amount of initial outer embryos in all simulations corresponding to each scenario. In Fig.~\ref{fig:remocion}, we present the averaged percentages of removed outer embryos with error bars for the six different scenarios. From this, we can observe that the percentage of removed outer embryos is an increasing function of the giant planet's mass. In fact, while a giant planet of of 1 $\textrm{M}_{\textrm{sat}}$ removed $\sim$ 40\% of the outer embryos, the massive giants of 2 $\textrm{M}_{\textrm{jup}}$ and 3 $\textrm{M}_{\textrm{jup}}$ removed $\sim$ 60\% and $\sim$ 70\% of them, respectively.

\begin{figure}
\includegraphics[width=0.45\textwidth]{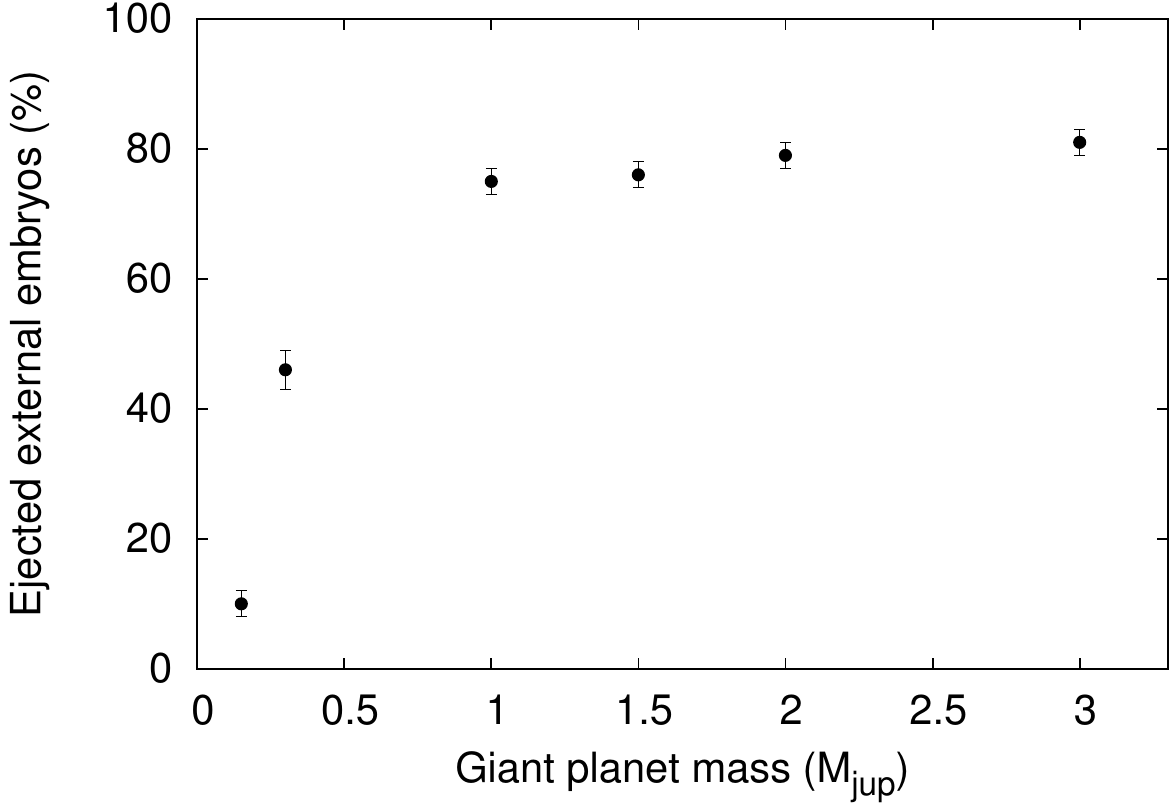}
\caption[Ejected outer embryos]{Average percentages of the amount of ejected outer embryos over the total amount of the removed outer embryos in each work scenario with their associated error.}
\label{fig:eyecciones}
\end{figure}

Once it is made, we calculated the percentage of outer embryos ejected from the system in each of our scenarios of study. In the $N$-body simulations, one body was assumed to be ejected if it reached a distance from the central star greater than 100 au. For each scenario, we computed the number of ejected outer embryos in every simulation. Then, we calculated the averaged percentages of ejections adding the number of ejected outer embryos in each simulation and normalizing it to the total amount of removed external embryos in all simulations corresponding to each scenario. In Fig.~\ref{fig:eyecciones}, we show these average percentages with their associated error bars. Our results suggest that the percentage of ejected outer embryos is an increasing function of the giant's mass. However, the reader can see that the averaged percentage of ejected outer embryos does not show significant changes for giant planets with masses between 1 $\textrm{M}_{\textrm{jup}}$ and 3 $\textrm{M}_{\textrm{jup}}$. The rest of the removed outer embryos that were not ejected from the system ended being accreted by other bodies of the system or hitting with the central star.

According to that presented in Fig.~\ref{fig:eyecciones}, the ejection of embryos from the system is a very efficient process in those work scenarios that host a giant planet more massive than 1 $\textrm{M}_{\textrm{jup}}$. To a better understanding, we calculate the heliocentric escape velocity of a small body located at 3 au. We realized that less massive giant planets (0.5 $\textrm{M}_{\textrm{sat}}$ and 1 $\textrm{M}_{\textrm{sat}}$) can not disperse small bodies with greater velocity than the heliocentric escape velocity at 3 $\textrm{au}$, while more massive giant planets are able to disperse small bodies with greater velocities as well, which favors their ejection from the system. We considered that our systems of study may significantly contribute to the production of free-floating rocky planets. A detailed study about it is out of the scope of the present research. 

\begin{figure}
\includegraphics[width=0.45\textwidth]{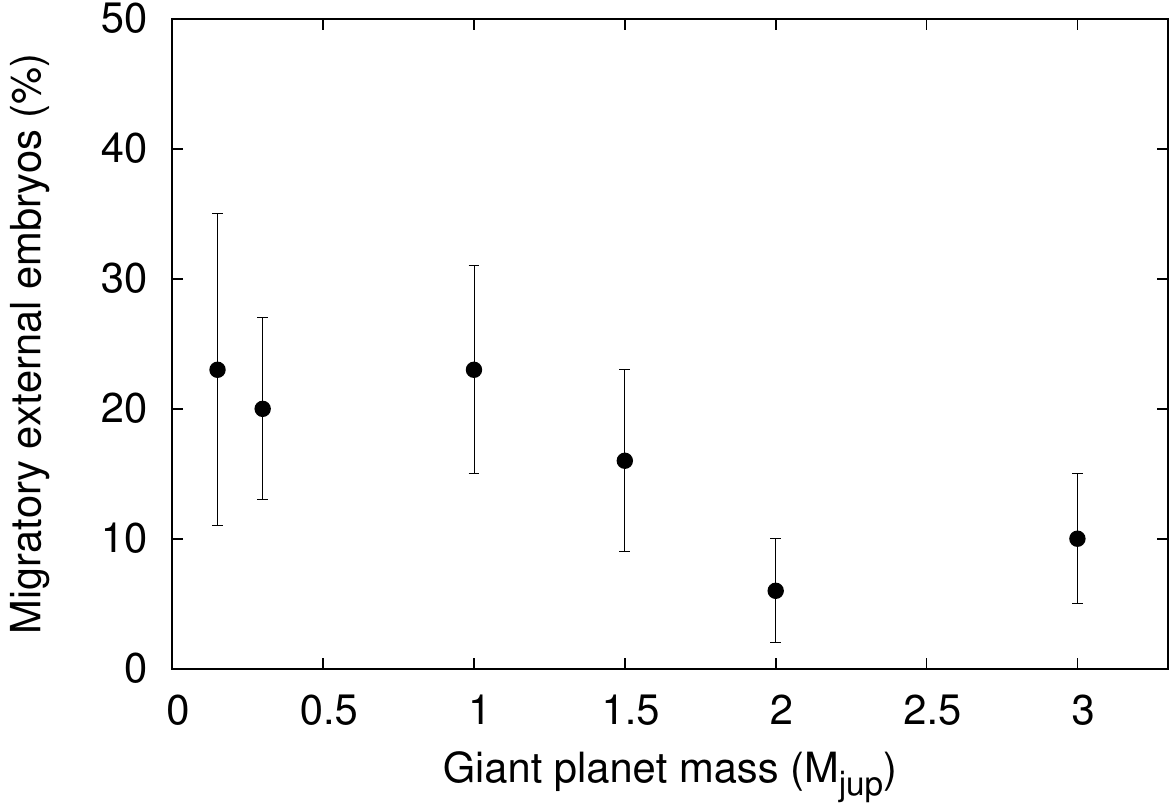}
\caption[Migrated outer embryos]{Average percentages of the amount of migrated outer embryos in each work scenario with their associated error.}
\label{fig:migraciones}
\end{figure}

An interesting analysis is that concerning the percentage of outer embryos that moved inward and ended in the inner region with a semi-major axis $a \leq$ 2.5 au in each work scenario. It is worth remarking that if the planet-planet scattering of outer embryos takes place in a given simulation, a single outer embryo moves inward reaching the inner region with $a \leq$ 2.5 au. In this context, we added those migrated embryos over the total amount of simulations per scenario. In Figure~\ref{fig:migraciones}, we showed the averaged percentage of outer embryos that moves inward with the corresponding error bars for each work scenario. It could be noticed that giant planets with masses less than 1 $\textrm{M}_{\textrm{jup}}$ allow the passage of outer embryos to the inner system up to $25\%$,  while giant planets of 2 $\textrm{M}_{\textrm{jup}}$ and 3 $\textrm{M}_{\textrm{jup}}$ allow less than $10\%$ of them. Moreover, we could see that a giant planet of 1 $\textrm{M}_{\textrm{jup}}$ represents a limit from which the amount of outer embryos which move inward the inner system, starts to decrease. This result is important because it shows the behavior of massive giant planets as barriers to the passage of water-rich material into the inner system. It might be of astrobiological interest, since it could allow us to set restrictions in observations, discarding those systems hosting a giant planet more massive than a few Jupiter masses to act as a barrier, decreasing the probabilities to find planets with water content in the inner regions of the system.

\begin{figure}
\includegraphics[width=0.46\textwidth]{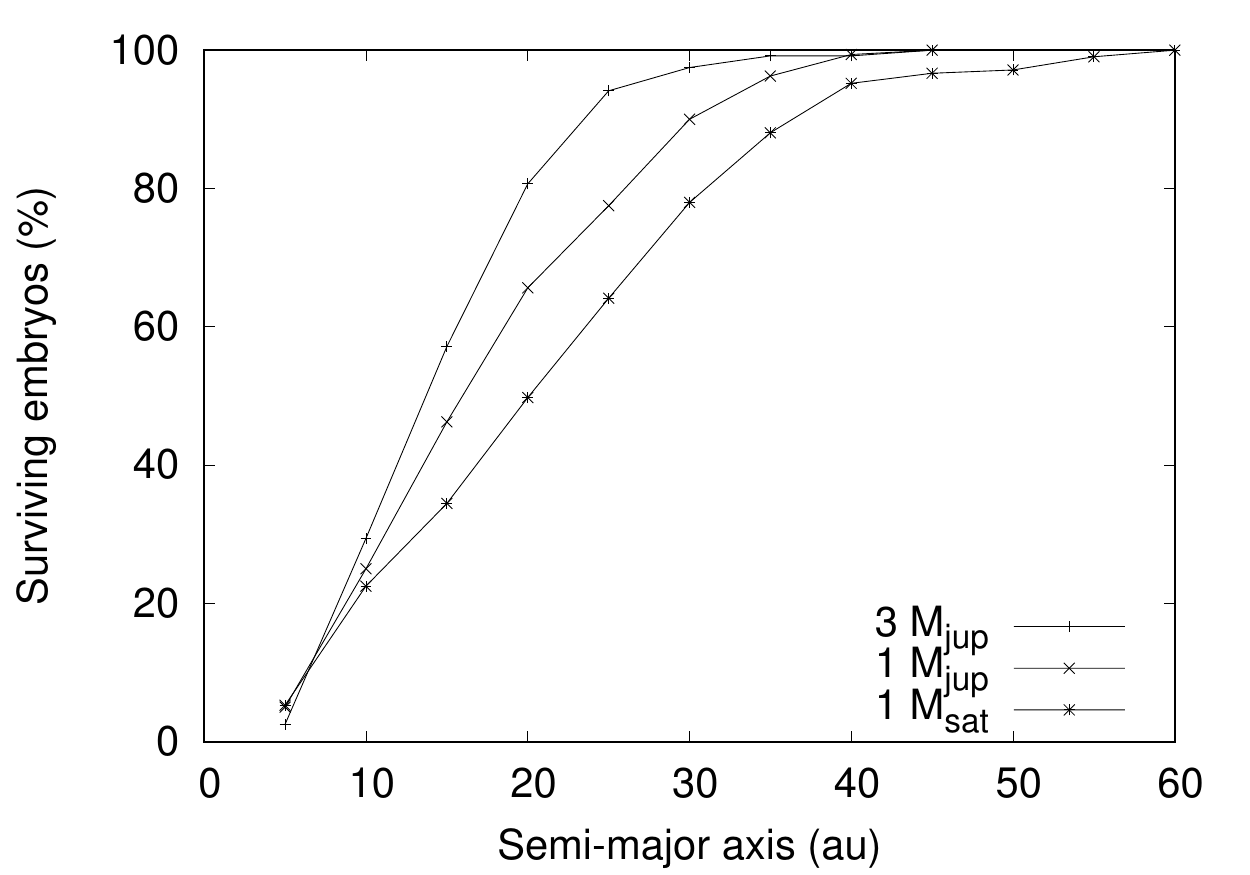}
\caption[Dispersion curves]{Cumulative percentages of the amount of surviving embryos in scenarios 2, 3 and 6, as a function of the semi-major axis.}
\label{fig:dispersion}
\end{figure}

In order to analyze the resulting system structure, we studied the semi-major axis distribution of the surviving embryos for each scenario. In Fig.~\ref{fig:dispersion} we represent the cumulative percentage of those surviving embryos as a function of the semi-major axis for three different scenarios: 3 $\textrm{M}_{\textrm{jup}}$, 1 $\textrm{M}_{\textrm{jup}}$ and 1 $\textrm{M}_{\textrm{sat}}$. We note from the figure that systems with less massive giant planets present the most extended systems. In fact, on the one hand, systems with a giant of 3 $\textrm{M}_{\textrm{jup}}$ presents more than $95\%$ of embryos inside $a \leq 30$ \textrm{au}, while systems with a giant of 1 $\textrm{M}_{\textrm{sat}}$ have less than $80\%$ of embryos up to that distance, reaching values of $a = 50$ \textrm{au}.

From the study of the evolution of planetesimals, we could say that in all of our scenarios, more than $90\%$ of them were removed. Furthermore, we found that most of the removed ones were ejected of the system leaving a small amount of available planetesimals to be accreted by other bodies of the system.

\subsection{Survival of planets in the HZ}

\begin{figure*}
\includegraphics[width=0.45\textwidth]{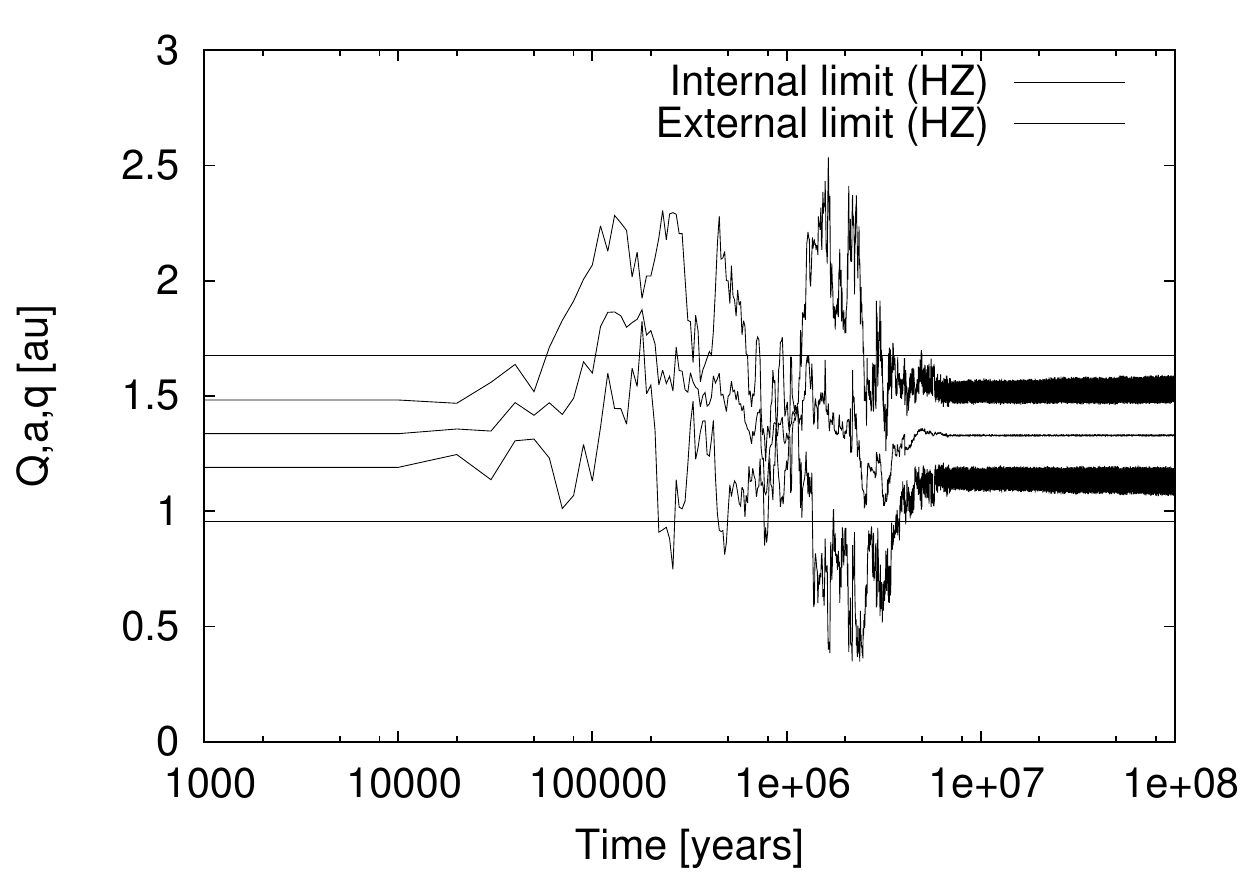}
\includegraphics[width=0.46\textwidth]{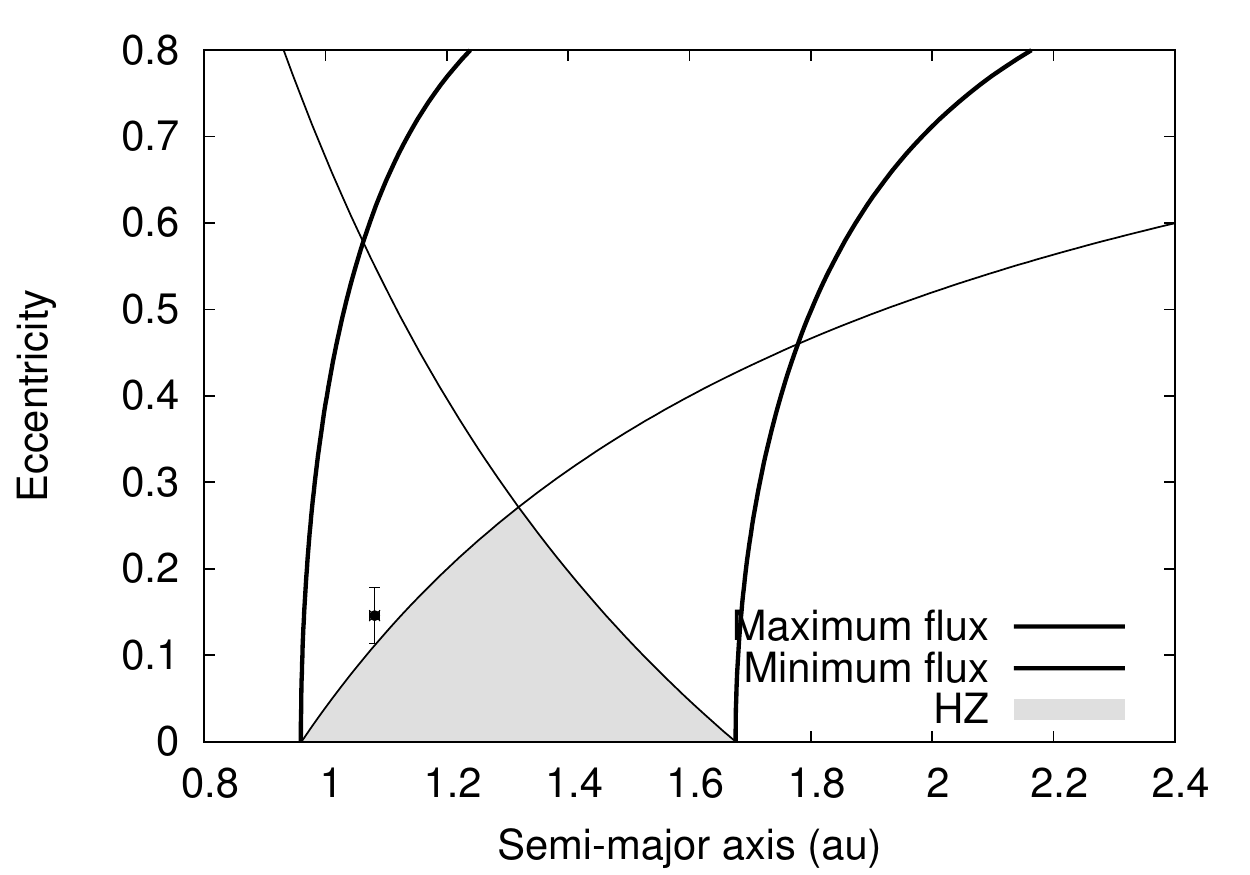}
\caption[Planets in habitable zone]{On the left, we could see, from top to bottom, the temporal evolution of aphelion Q distance, semi-major axis a, and perihelion q distance, which are completely contained into the limits of the HZ, of a planet from scenario 6. On the right, we present the averaged values of semi-major axis and eccentricity, with their associated change bars, of a given planet from scenario 6. The gray shaded area represents the conservative HZ, locked in between constant perihelion and aphelion curves. Thicker black curves represent maximum and minimum averaged flux.}
\label{fig:criterio}
\end{figure*}

After studying how a single giant planet located around the snow line affects the embryo evolution in a given system once the gas has dissipated, we analyze how many surviving embryos became planets in the HZ. In particular, we are interested in describing the physical properties of the planets that survive in the HZ in each system of work, focusing in their masses and water contents. To do this, we use the definition of HZ proposed in Sect.~\ref{sec:hzmodel}. As we have already mentioned, we consider that a planet is potentially habitable if its semi-major axis $a$, and its perihelion $q$ and aphelion $Q$ distances are within the limits of the HZ. However, if the planet's orbit is not fully contained in the HZ but the perihelion or the aphelion is near the limits of such region, we use the averaged flux criterion proposed by \citet{Williams2002} in order to analyze its habitability. Figure~\ref{fig:criterio} shows both of such examples. On the one hand, in the left panel we can see the temporal evolution of the aphelion distance, semi-major axis and perihelion distance of a planet from scenario 6, respectively. From the beginning, such a planet evolves showing significant changes in its semi-major axis and eccentricity. However, after a few Myr, the planet survives evolving with its orbit fully contained within the HZ conservative limits up to the end of the integration, which are illustrated by the horizontal black lines. On the other hand, the right panel shows the averaged values of the semi-major axis and eccentricity, together with their respective variation bars, of a given planet from scenario 6. The gray shaded area represents the conservative HZ, which is locked in between constant perihelion and aphelion curves. The change bars of the semi-major axis and eccentricity allow us to infer that the planet's orbit is not fully contained in the conservative HZ since the perihelion distance is slightly smaller than it is the inner limit during almost the whole integration. However, planet's orbit is fully contained within the region delimited by the minimum and maximum allowed stellar flux curves (thicker black curves). According to this, the planet under study is considered habitable in the present research from the averaged flux criterion. 
Taking into account all the simulations made for all our work scenarios, we found that $37 \%$ of the potentially habitable planets have their orbits completely contained into the limits of the HZ, while the remaining $63 \%$ have been selected by the averaged flux criterion. We must mention that all of those planets presents eccentricities $e < 0.6$, in order to apply the averaged flux criterion on them. A planet with an eccentricity higher than $0.6$ cannot sustain surface liquid water during the whole orbital period \citep{Bolmont2016}.
\begin{figure*}
\includegraphics[width=0.95\textwidth]{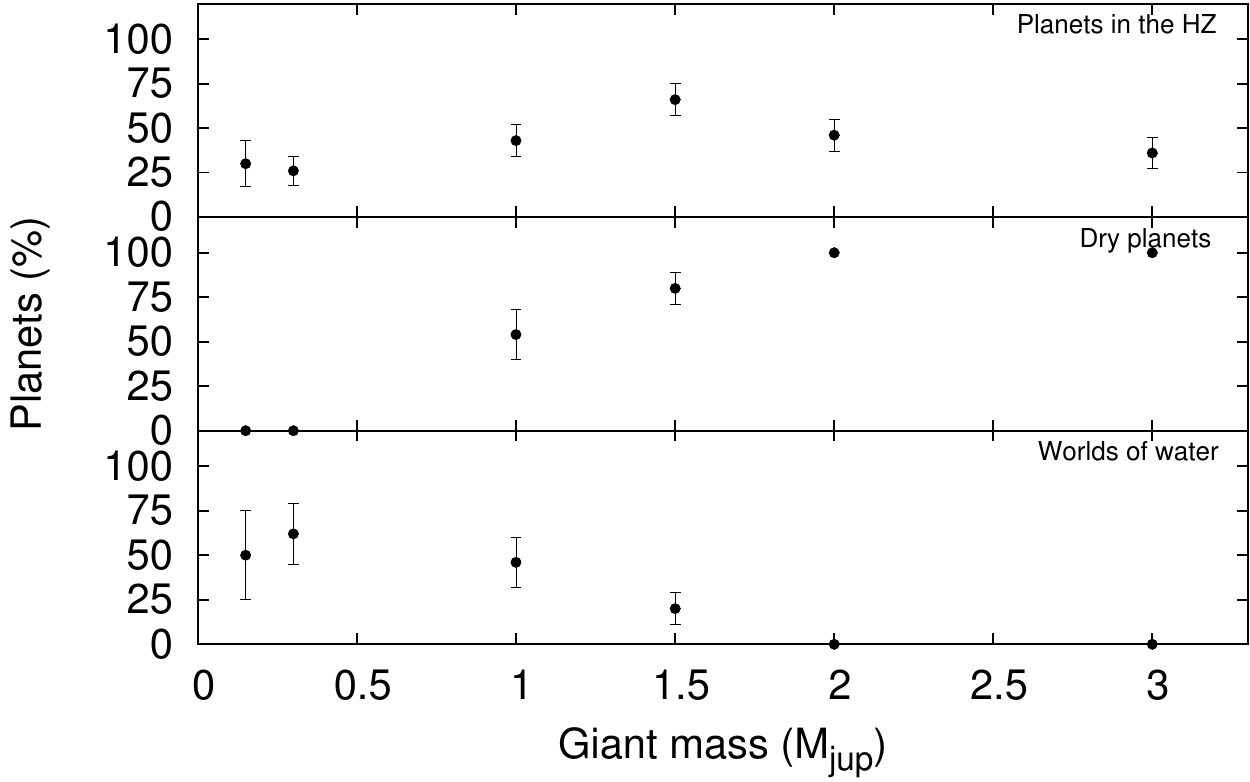}
\caption[Planets]{Average percentages of the total amount of planets found in the HZ in each work scenario (top panel), dry planets (middle panel) and water worlds (bottom planets). The last two computed over the total amount of planets found in the HZ.}
\label{fig:planetaszh}
\end{figure*}

It is worth remarking that the six work scenarios of the present research produced planets in the HZ. In each scenario, some of the $N$-body simulations formed only one HZ planet, while the other ones did not produce anyone. To get a better understanding about the formation efficiency of potentially habitable planets in each work scenario, we computed the averaged percentages of HZ terrestrial-like planets over the total of simulations carried out per scenario, which can be seen in the top panel of Fig.~\ref{fig:planetaszh}. From this, the percentage of potentially habitable planets formed in the six work scenarios ranges from 25\% to 75\%. Notice that the scenario 4 represents the most efficient one, producing 20 potentially habitable planets over 30 $N$-body simulations.

In general terms, our simulations produced two different classes of planets in the HZ: {\it dry worlds} and {\it water worlds}. On the one hand, the dry worlds evolved from accretion seeds\footnote{Following \citet{Raymond2009}, we define a planet accretion seed as the largest embryo involved in its collisional history.} that were located inside the snow line at the end of the gas phase and did not accrete any water during their collisional history. On the other hand, the water worlds survived in the HZ, which evolved from accretion seeds initially located in a water-rich region beyond the snow line, present between $28 \%$ and $50 \%$ of water by mass. It is worth noting that $12 \%$ of  water worlds that survived in the HZ evolved from accretion seeds initially located inside the snow line but received an impact from a water-rich outer embryo, which allows us to understand their very high water contents at the end of the simulations, between $22 \%$ and $28 \%$ of water in mass.

The percentage of dry and water worlds formed in each of our work scenarios is illustrated in the middle and bottom panels of Fig.~\ref{fig:planetaszh}, respectively. From this, it is possible to observe that the percentage of dry (water) worlds that survive in the HZ in each work scenario increases (decreases) as a function of the giant planet's mass. In this context, the scenarios 5 and 6, represent extreme cases since all planets that survived in the HZ at the end of the integration are dry worlds. In fact, these scenarios did not form water worlds in the total sample of simulations.   

The scenarios 1 and 2 also represent particular cases. On the one hand, they did not form dry worlds in the HZ and a high percentage of the planets that survived in the HZ are water worlds. On the other hand, it is important to mention that the scenario 2 represents the only one that produced planets in the HZ with small percentages of water by mass, because they received, at least, one impact from planetesimals rich in water, in addition to dry or water worlds. They represent the $37 \%$ of potentially habitable planets and present less than $2.5 \%$ of water by mass. Moreover, the scenario 1 is the only one that presents the own giant into the HZ at the end of $50 \%$ of the simulations. The remaining $50 \%$ of planets in the HZ correspond to water worlds.

As we described above, in all scenarios, different kind of planets were formed in the HZ with respect to their amount of water by mass. From these results, we can also say that they differ in the value of their masses. In scenarios 1 and 2 no sub-Earth planet was formed. Planets in the HZ reached masses up to $8$ $\textrm{M}_{\textrm{\earth}}$ (without counting the giant itself which migrated in scenario 1). All the others work scenarios, formed such as sub-Earth planets as super-Earth planets. The Scenario 3 formed the most massive super-Earth planet of $11.16$ $\textrm{M}_{\textrm{\earth}}$ which is also a water world. scenarios 5 and 6 only reached masses up to $5$ $\textrm{M}_{\textrm{\earth}}$. They could also form Earth-like planets of a mass of $0.98$ $\textrm{M}_{\textrm{\earth}}$, which were dry worlds.

The bottom panel of Fig.~\ref{fig:planetaszh} allows us to infer a very interest result concerning the role of a giant planet as dynamical barrier in the evolution of terrestrial-like planets and water delivery in the HZ of the system. In fact, our results suggest that a single gaseous giant of 1 $\textrm{M}_{\textrm{jup}}$ located around the snow line seems to represent a limit mass above which the efficiency of formation of water worlds in the HZ significantly decreases. This result is relevant since it allows us to define a selection criterion for the search of potentially habitable exoplanets in systems that host a single giant planet close to the snow line around solar-type stars.


\section{Conclusions and discussions}

In the present work, we analyzed how a single giant planet located around the snow line affects the dynamical evolution of terrestrial-like planets and water delivery in the HZ after the gas dissipation in solar-type star systems. Our study showed a statistical analysis based on results obtained from $N$-body simulations of planetary accretion. In order to analyze the sensitivity of our analysis regarding the mass of giant planets, we carried out $N$-body simulations for six different work scenarios, in which the mass of the giant planet was varied between 0.5 $\textrm{M}_{\textrm{sat}}$ and 3 $\textrm{M}_{\textrm{jup}}$.   

Our results suggest that a Jupiter-mass planet could represent a limit mass above which the amount of water-rich embryos that moves inward from beyond the snow line starts to decrease. From this, a giant planet more massive than one Jupiter-mass might results to be an efficient dynamical barrier to inward-migrating water-rich embryos. 

This result has relevant implications concerning the survival of water-rich terrestrial planets in the HZ of a given system. In fact, while the six work scenarios of our research produced planets in the HZ, the percentage of dry (water) worlds that survive in the HZ increases (decreases) as a function of the giant's mass. In this context, a Jupiter-mass planet located around the snow line seems to represent a limit mass above which the number of water worlds in the HZ significantly decreases. In fact, those scenarios that host a perturbing of 2 $\textrm{M}_{\textrm{jup}}$ and 3 $\textrm{M}_{\textrm{jup}}$ around the snow line represent extreme cases, which did not produce water worlds in the HZ in any simulation.

It is important to remark that the results previously described should be interpreted in the context of the numerical model used to carry out the $N$-body simulations. In fact, the \textsc{MERCURY} code used in the present study treats all collisions as inelastic mergers, which conserve the total mass and the water content of the interacting bodies. Thus, the masses and water contents of all planets formed in HZ should be interpreted as upper limits. 

Recent investigations based on hydro dynamical simulations have shown that collisions are not always perfect mergers. In fact, studies such as those developed by \citet{Leinhardt2012} and \citet{Genda2012} analyze the limits of the different collisional regimes and describe the size and velocity distribution of the post-collision bodies. Later, \citet{Chambers2013} used the results of those works to carry out $N$-body simulations of terrestrial planet formation incorporating fragmentation and hit-and-run collisions. In such a work, the author compared those $N$-body simulations with other ones previously developed assuming all collisions as perfect mergers. The general results derived by \citet{Chambers2013} suggested that the final planetary
systems produced in the two numerical models were similar. However, the author observed that planets that result in a given system have somewhat smaller masses and eccentricities when a more realistic treatment is included in the model. Recently, \citet{Quintana2016} studied giant impacts on Earth-like planets in the last stage of the evolution of a planetary system, using $N$-body simulations, which included fragmentation and hit-and-run collisions.

On the other hand, \citet{Dvorak2015} developed hydrodynamic simulations to infer the amount of water in fragments after a collision for different velocities and impact angles. They found that most of the water is retained by the survivor body for impact angles $\alpha \lse 20^{\circ}$ and velocities $\nu \lse 1.3 \nu_{esc}$, with $\nu_{esc}$ their escape velocity. As a last work, \citet{Mustill2017} explored the effects of implementing a more realistic collision treatment on in-situ formation of planets which radial distances of few tenth of au. Taking those results into account, we consider that is important to include a more realistic treatment of the collisions and the evolution of water in the $N$-body code, in order to refine our percentages of water in the final potentially habitable planets found in all the work scenarios and verify if the dry planets that we found were totally dry or if they could present a small amount of water in mass.

One last thing to take into account, is the fact that we fix the snow line in 2.7 au. We are aware of the evolution of the snow line with time and its profile according to a Sun-like star \citep{Ciesla2015}. However, we consider a fix snow line as a good approximation during our integration time and a distance of separation between dry and water rich material at the beginning of our simulations, as it is assumed by different authors such as \citet{Raymond2004}, \citet{Obrien2006}, \citet{Raymond2009}, \citet{Ronco2014}, \citet{Zain2018}, among others, who worked with $N$-body simulations in the last stage of the formation of a planetary system around a solar-mass star, once the gas has been already dissipated from the disk. Even though, we consider that it would be a good experiment to move the snow line inward as \citet{Ciesla2015} in their simulations in order to test the sensitivity of our results with respect to an inner separation between dry and water-rich material. This could have important consequences with respect to the final amount of water in mass of the resulting final planets in the HZ. However, this analysis is out of the scope of this work.

We consider that the present work allows us to get a better understanding of the role of giant planets in the formation of terrestrial planets around a Sun-like star. We infer that our results could give a selection criteria for future searches of potentially habitable exoplanets.

\section*{Acknowledgments}

This work was partially financed by CONICET through the PIP 0436/13. Moreover, the authors acknowledge the financial support by FCAGLP for extensive use of its computing facilities.





\bibliographystyle{mnras}
\bibliography{msanchezfinal} 








\bsp	
\label{lastpage}
\end{document}